\documentclass[prl,a4paper,aps,twocolumn,showpacs]{revtex4-1}

\usepackage{amsfonts,amsmath,bm}
\usepackage{graphicx}
\newcommand{\rl}{\rangle\!\langle}
\newcommand{\kk}{\bm{k}}
\newcommand{\rr}{\bm{r}}
\newcommand{\sumkl}{\sum_{\bm{k}\lambda}}
\newcommand{\wk}{\omega_{k}}
\newcommand{\bkl}{b_{\bm{k}\lambda}}
\newcommand{\bkld}{b_{\bm{k}\lambda}^{\dag}}

\begin{document}

\title{Enhanced spontaneous emission from inhomogeneous
  ensembles of quantum dots is induced by short-range couplings} 

\author{Micha{\l} Kozub}
\author{{\L}ukasz Pawicki}
\author{Pawe{\l} Machnikowski}
 \email{Pawel.Machnikowski@pwr.wroc.pl} 
\affiliation{Institute of Physics, Wroc{\l}aw University of
Technology, 50-370 Wroc{\l}aw, Poland}

\begin{abstract}
We study theoretically the spontaneous emission from an
inhomogeneous ensemble of quantum dots in the weak excitation
limit. We show that collective, superradiance-like 
effects lead to an enhanced emission rate in the presence of
sufficiently strong coupling between the dots in agreement with
experimental observations, which means that the
quantum dot
sample cannot be treated as an ensemble of individual emitters. We
demonstrate also that the collective behavior of the quantum dot
system relies on short-range interactions, while long-range dipole
couplings are too weak to 
have any impact on the emission dynamics for a system with a realistic
degree of inhomogeneity.
\end{abstract}


\maketitle

%\section{Introduction} 

The unique atomic-like optical properties of
semiconductor quantum dots (QDs) allow one to transfer
ideas and technologies from atomic 
systems to solid state structures and devices. 
Such quantum optics concepts successfully
implemented in QD systems range from laboratory-scale coherent optical
control 
 experiments \cite{zrenner02} to commercially available devices, like QD
lasers \cite{kirstaedter94}. While similarities
between atoms and QDs may be very
useful, these two systems differ essentially with some
respects. One of these differences is the inhomogeneity of the QD
characteristics due to inevitable randomness of the physical
properties of these
structures. This becomes important when collective
interaction of QDs with their environment is essential, e.g.,
in lasers or some quantum information devices \cite{zanardi99a}. Understanding
the collective evolution of non-identical, coupled emitters can also
be beneficial to the current study of other semiconductor \cite{cesar11}, 
plasmonic \cite{teperik12}, hybrid \cite{haridas11} or
atomic \cite{pritchard12} systems where a similar interplay of
collectivity, inhomogeneity, and  
interactions plays a crucial role.

In an experiment performed on a dense ensemble of CdSe QDs
\cite{scheibner07}, the decay of luminescence was shown to
accelerate as the number of emitting dots
increased. This means that, in their interaction with the radiation
field, the QDs in the ensemble are not independent
objects. The existence of such a cooperative effect for QDs is
remarkable as the ensemble broadening of the transition
energies is 3-4 orders of magnitude larger than the
radiative broadening of the QD emission line. The experimentally
observed effect was quite strong 
already for fewer than $100$ QDs so that an accidental
spectral overlap  between two or more dots was clearly very unlikely. The
cooperative effect was therefore attributed to the existence of
long-range (LR) coupling between the dots. 

Subsequent theoretical analysis
\cite{sitek07a,sitek09b,averkiev09} confirmed that coupling between 
non-identical QD 
emitters stabilizes the collective effects in the spontaneous
emission. However, in order to overcome the
inhomogeneity the coupling 
must be at least comparable to the transition energy mismatch between
the emitters \cite{sitek07a,sitek09b}, while
 the LR dipole (F\"orster) couplings do not exceed a
fraction of meV even for dots separated be a few nanometer distance
\cite{lovett03b} and drop down to about 1~$\mu$eV for inter-dot
separations around 30~nm, which is the average value for the
experimentally studied sample \cite{scheibner07}. These values are
at least two orders of magnitude lower than the transition energy
inhomogeneity which is on the order of a few tens of meV. Thus, 
the LR dipole couplings seem unlikely to underlie
the enhanced emission. It is known, however, that also other couplings
of different nature exist in QD ensembles
\cite{kazimierczuk09,kasprzak10}. 

In this paper, we
present the results of theoretical modeling of the
spontaneous emission from an inhomogeneous ensemble of QDs
and clarify the origin of the observed enhanced spontaneous
emission.  We extend the standard model of identical emitters 
\cite{gross82,yukalov10} by including ensemble
inhomogeneity and simulate the 
evolution of a planar ensemble of non-identical, randomly distributed
QDs in the weak excitation regime. 
Our analysis confirms that
the cooperative interaction of the QDs with the surrounding
electromagnetic (EM) vacuum can lead to an increased spontaneous emission
rate. We show, however, that the dipole coupling between the QD
emitters is insufficient for the appearance of cooperative behavior of these
strongly inhomogeneous systems. Enhanced emission appears only if one
includes short-range (SR) couplings between the dots, which may be due to a
combination of tunneling (wave function overlap between the
neighboring QDs) and Coulomb correlations.  

We model a system of $N$ QDs located at points $\bm{r}_{\alpha}$.
The typical energy distance to p-shell exciton states in CdSe/ZnSe
dots is at about 100~meV \cite{sotier09}, much larger than
the inhomogeneous broadening of the s-shell states, which allows us to 
restrict the discussion to the fundamental transition. In addition, we fix
the polarization. Hence each of the dots is represented as a two-level
system with the basis states $|0\rangle_{\alpha}$ and
$|1\rangle_{\alpha}$, where $\alpha$ labels the dots. The fundamental transition
energy in the dot $\alpha$ is $E_{\alpha}$. The dots are coupled by
SR couplings described effectively by the coupling
potentials
$\Omega_{\alpha\beta}^{(\mathrm{T})}$.
The Hamiltonian for
the EM interactions is transformed to the dipole form using
the Power--Zienau--Wooley transformation \cite{cohen89} and reads
\begin{eqnarray*}
H & = & \sum_{\alpha=1}^{N}E_{\alpha}\sigma_{\alpha}^{\dag}\sigma_{\alpha}
+\sum_{\alpha\neq\beta}\Omega_{\alpha\beta}^{(\mathrm{T})}
\sigma_{\alpha}^{\dag}\sigma_{\beta}
+\sumkl\hbar\wk \bkld\bkl
 \\
&&-\frac{1}{\varepsilon_{0}\varepsilon_{\mathrm{r}}}
\sum_{\alpha=1}^{N}
(\bm{d}_{0}\sigma_{\alpha}+\bm{d}_{0}^{*}\sigma_{\alpha}^{\dag}) \cdot
\bm{D}(\bm{r_{\alpha}}),
\end{eqnarray*}
where the EM field is represented by 
the displacement operator
\begin{displaymath}
\bm{D}(\bm{r})=i\sumkl \sqrt{\frac{
\hbar\varepsilon_{0}\varepsilon_{\mathrm{r}}\wk}{2V}}
\hat{e}_{\bm{k}\lambda}\bkl e^{i\kk\cdot\rr} +\mathrm{h.c.}
\end{displaymath}
Here $\bkl$ is the annihilation operator for a photon with wave
vector $\kk$ and polarization $\lambda$, $\wk$ is the photon frequency,
$\hat{e}_{\bm{k}\lambda}$ is the unit polarization vector, 
$\sigma_{\alpha}=(|0\rl 1|)_{\alpha}$ is the transition operator
for the dot $\alpha$, $\bm{d}_{0}$ is the interband
matrix element of the dipole moment (we assume that all the interband
dipoles are identical), $c$ is the 
speed of light in vacuum, $\varepsilon_{0}$ is the 
vacuum permittivity and $\varepsilon_{\mathrm{r}}$ is
the dielectric constant of the semiconductor medium.

The evolution equation for any electronic operator $Q$
can be obtained along the lines worked out in Ref.~\cite{lehmberg70a}:
One writes down the equations of motion for the 
electronic and photonic operators, eliminates the latter, neglects the
off-resonant terms and radiation-induced energy shifts, and performs the Markov
approximation. It is assumed that no
external fields are present.
The resulting evolution equation for the average value of the
electronic operator $Q$ has the form
\begin{eqnarray}
\langle\dot{Q}\rangle & = & \frac{i}{\hbar}\sum_{\alpha}\epsilon_{\alpha}
\left\langle\left[ \sigma_{\alpha}^{\dag}\sigma_{\alpha},Q  \right]\right\rangle
+\sum_{\alpha\neq\beta}i\Omega_{\alpha\beta}
\left\langle\left[ 
\sigma_{\alpha}^{\dag} \sigma_{\beta},Q\right]\right\rangle \nonumber\\
&&+\sum_{\alpha,\beta}\Gamma_{\alpha\beta}
\left\langle\sigma_{\alpha}^{\dag}Q\sigma_{\beta}
-\frac{1}{2}\left\{\sigma_{\alpha}^{\dag} \sigma_{\beta},Q
  \right\} \right\rangle,
\label{eq-motion}
\end{eqnarray}
where
$\Omega_{\alpha\beta}=\Omega_{\alpha\beta}^{(\mathrm{rad})}
+\Omega_{\alpha\beta}^{(\mathrm{T})}$ and
\begin{equation}
\Omega_{\alpha\beta}^{(\mathrm{rad})} =\Gamma
G_{\alpha\beta}(k_{0}r_{\alpha\beta}),
\quad
\Gamma_{\alpha\beta} =
\Gamma F_{\alpha\beta}(k_{0}r_{\alpha\beta}).
\label{Om-Gamma}
\end{equation}
Here $\bm{r}_{\alpha\beta}=\bm{r}_{\beta}-\bm{r}_{\alpha}$,
$\Gamma=
|d_{0}|^{2}k_{0}^{3}/(3\pi\varepsilon_{0}\varepsilon_{\mathrm{r}}\hbar)$
is the spontaneous emission (radiative recombination) rate for a
single, isolated QD,
$k_{0}=nE/(\hbar c)$ is the average resonant wave vector in the
dielectric medium with the refractive index $n$, 
\begin{eqnarray*}
G_{\alpha\beta}(\xi) & =& 
\frac{3 -9 | \hat{d}_{\|} |^{2}}{4} 
\left(\frac{\sin\xi}{\xi^{2}} + \frac{\cos\xi}{\xi^{3}}  \right)
-\frac{3| \hat{d}_{\bot} |^{2}}{4} \frac{\cos\xi}{\xi},\\
F_{\alpha\beta}(\xi) & = &
\frac{3 -9 | \hat{d}_{\|} |^{2}}{2}
\left(\frac{\cos\xi}{\xi^{2}} -\frac{\sin\xi}{\xi^{3}}  \right)
+\frac{3 |\hat{d}_{\bot} |^{2}}{2} \frac{\sin\xi}{\xi},
\end{eqnarray*}
and $\hat{d}_{\|}$ and $\hat{d}_{\bot}$ are the
components of $\hat{\bm{d}}_{0}=\bm{d}_{0}/d_{0}$ parallel and
perpendicular to  $\hat{\bm{r}}_{\alpha\beta}$, respectively.
The first expression in Eq.~\eqref{Om-Gamma} describes the dipole couplings
between the interband dipoles confined in the QDs (referred to as
\textit{F\"orster coupling} or \textit{dispersion forces}).
We assume that only heavy hole
excitons are involved, so that $\bm{d}_{0}=d_{0}(1,\pm i,0)^{T}/\sqrt{2}$,
hence $|\hat{d}_{\|}|^{2}=|\hat{d}_{\bot}|^{2}=1/2$ since the QDs
are distributed in the $xy$ plane. An admixture of light hole states
would 
reduce the dipole coupling at short distances. However, typical 
admixture of at most several per cent
\cite{koudinov04} will only bring a small quantitative
correction to our results and will not affect the qualitative
conclusions of this paper. 

In order to find the evolution of the total number of excitons, 
$N_{X}=\sum_{\alpha}\langle\sigma_{\alpha}^{\dag}\sigma_{\alpha}\rangle$,
we use Eq.~\eqref{eq-motion} to find the equations of motion for the
quantities
$x_{\alpha\beta}=\langle\sigma_{\alpha}^{\dag}\sigma_{\beta}\rangle$. These
quantities are dynamically coupled to higher order terms
of the form 
$\langle\sigma_{\alpha}^{\dag}\sigma_{\beta}^{\dag}\sigma_{\gamma}\sigma_{\delta}\rangle$.
In general, in an inhomogeneous ensemble, there are no
constants of motion in the $N$-dot dynamics and the system is
described by the $2^{2N}$ elements of the general density matrix,
which is beyond the simulation capabilities already for several
dots. Here, we will restrict the discussion to the low excitation
case, with at most one exciton present in the QD ensemble (in this
case, collectivity results from delocalization of this single
excitation over many emitters jointly interacting with the field). Then, the
normally ordered higher order terms vanish and one obtains a closed
system of equations 
of motion in the form
\begin{eqnarray}
\dot{x}_{\alpha\beta} & = & 
\frac{i}{\hbar}\left( \epsilon_{\alpha}-\epsilon_{\beta} \right) x_{\alpha\beta}
+i\sum_{\gamma}\left( \Omega_{\gamma\alpha}x_{\gamma\beta}
-\Omega_{\beta\gamma}x_{\alpha\gamma}\right) \nonumber\\
&&
- \frac{1}{2}\sum_{\gamma}\left( \Gamma_{\gamma\alpha}x_{\gamma\beta}
+\Gamma_{\beta\gamma}x_{\alpha\gamma}\right),
\label{evol}
\end{eqnarray}
where it is assumed that $\Omega_{\gamma\gamma}=0$.

The photon detection rate for a hypothetical ideal detector collecting from a
solid angle $d\Omega$ around a point $\bm{R}$ at a
large distance from the sample is 
proportional to the field correlation function \cite{walls08},
\begin{equation}\label{detection-D}
J(\hat{\bm{R}})d\Omega=
\frac{2cR^{2}}{nE\varepsilon_{0}\varepsilon_{\mathrm{r}}}
\left\langle \bm{D}^{(-)}(\bm{R},t) \cdot \bm{D}^{(+)}(\bm{R},t)
\right\rangle
d\Omega,
\end{equation}
where $\bm{D}^{(\pm)}$ are the positive and negative frequency parts
of the field. The field originates from the interband dipoles and can
be related to the transition operators $\sigma_{\alpha}$. In the
Markov limit one finds \cite{lehmberg70a}
\begin{equation}\label{detection-sigma}
J(\hat{\bm{R}}) = \frac{3\Gamma
(1-|\hat{\bm{d}}_{0}\cdot\hat{\bm{R}}|) }{8\pi}
\sum_{\alpha\beta} e^{ik\bm{r}_{\alpha\beta}\cdot\hat{\bm{R}}}
\left\langle \sigma_{\alpha}^{\dag}(t)\sigma_{\beta}(t) \right\rangle.
\end{equation}
For a spectrally resolved detection, one has to replace the field
operators in Eq.~\eqref{detection-D} by spectrally filtered operators 
$\tilde{\bm{D}}^{(\pm)}(t)=\int dt' f(t-t') \bm{D}^{(\pm)}(t')$, where
$f(t)$ is the inverse Fourier transform of the filter function
$f(\omega)$. This results in an analogous replacement for the
$\sigma_{\alpha},\sigma_{\beta}$ operators in
Eq.~\eqref{detection-sigma}. The two-time correlation functions are
calculated by means of Eq.~\eqref{eq-motion} and the quantum
regression theorem \cite{walls08}, starting from the equal time
average known from the solution to Eq.~\eqref{evol}. For a
milli-electron-volt filter width, the relevant time differences are on
the picosecond scale, much shorter than the exciton life time, so that
the dissipative part of Eq.~\eqref{eq-motion} is neglected.

To model the experimental situation \cite{scheibner07}, we
consider ensembles of quantum dots placed on
square mesas. The spectral properties of the dots are
modeled by a Gaussian distribution of their transition energies
with the mean $E$ and variance $\sigma^{2}$. The
dots are located at 
random in the sample plane, with the restriction that the inter-dot
distance cannot be smaller than $10$~nm. The mesas are ``cut out'' 
from a larger sample, so that the number of dots fluctuates. At the
initial time, the QD ensemble is 
supposed to be coherently excited into the delocalized state
$|\psi(0)\rangle =(1/N)^{1/2}\sum_{\alpha}\sigma_{\alpha}^{\dag}|0\rangle$,
where $|0\rangle$ is the system ground state. For each set of
parameters, the results are averaged over many
repetitions with different, randomly generated QD
distributions. In the simulations presented here, we
assume $1/\Gamma=390$~ps, $n=2.6$,
$E=2.59$~eV and the QD surface density $\nu=10^{11}$ QDs/cm$^{2}$.
For the SR coupling between the QDs we choose the simplest exponential model,
$\Omega_{\alpha\beta}^{(\mathrm{T})}
=V_{0}\exp[-r_{\alpha\beta}/r_{0}]$, with the
amplitude $V_{0}=5$~meV and the spatial range $r_{0}=15$~nm, chosen to
fit the experimental data. 

\begin{figure}[tb]
\begin{center}
\includegraphics[width=85mm]{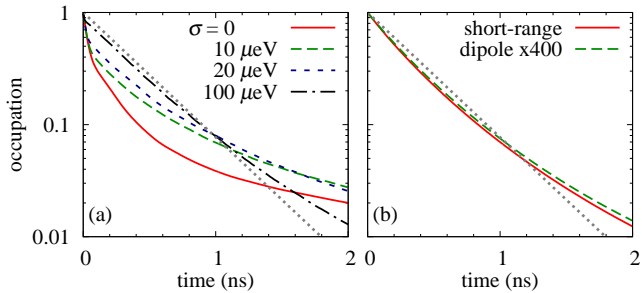}
\end{center}
\vspace{-5mm}\caption{\label{fig:inhom}The time dependence of
  the exciton occupation for a QD ensemble on a $0.03$~$\mu$m$^{2}$
  mesa (30 QDs on the average). (a) With LR dipole couplings
  only. Solid (red) line: homogeneous system. Dashed
  lines: three different inhomogeneous systems with the standard
  deviations of the fundamental transition energies $\sigma$ as
  indicated. (b) With additional SR couplings (solid red
  line) and with the dipole coupling artificially enhanced by a factor
of 400 (dashed green line) for $\sigma=28$~meV. In both figures, the
dotted (gray) line 
shows the exponential decay with the rate $\Gamma$ characteristic of a
single dot. The results are averaged over 20 random QD distributions (the
same for each case).}
\end{figure}

The decay of the exciton population in a system of 30 QDs on the
average with various degrees of inhomogeneity and coupled by
interactions of different kinds is shown in Fig.~\ref{fig:inhom}. If
only LR dipole couplings are included
[Fig.~\ref{fig:inhom}(a)] the evolution of a system of identical dots
is markedly non-exponential 
with a long stage of strongly accelerated emission (red solid
line). However, this effect vanishes very quickly as soon as the
inhomogeneity of the fundamental transition energies comes into
play. As can be seen in Fig.~\ref{fig:inhom} (black dash-dotted line),
already for $\sigma=0.1$~meV, the decay almost 
exactly follows the exponential evolution of a single dot, with only a
very short initial period of enhanced emission.

The results are completely different if a sufficiently strong
SR coupling is taken into account [Fig.~\ref{fig:inhom}(b),
red solid line]. 
Now, the exciton occupation decays with an enhanced
rate even for the realistic value of $\sigma=28$~meV \cite{scheibner07}. Moreover,
although the time dependence is not strictly exponential, the emission rate is
much more  constant over the two decades of occupation change shown. 

Although one could expect that the LR dipole interaction should
lead to qualitatively different dynamics than the SR coupling this
turns out not to be the case for the inhomogeneous QD system. As shown
in Fig.~\ref{fig:inhom}(b) (green dashed line), the evolution of the
exciton occupation 
in the hypothetical case of dipole couplings magnified by a factor of
400 is almost identical to that observed for
SR couplings for the parameters of these couplings chosen
here. The reason is that in any case the coupling is important for the
spontaneous emission only when its magnitude is at least comparable to
the energy mismatch between the emitters \cite{sitek07a}. For a
realistic, rather strongly inhomogeneous ensemble of QDs, only
coupling between relatively close QDs is strong enough to have any impact 
on the system dynamics. Hence, the properties of the interactions on
longer distances are irrelevant.

\begin{figure}[tb]
\begin{center}
\includegraphics[width=85mm]{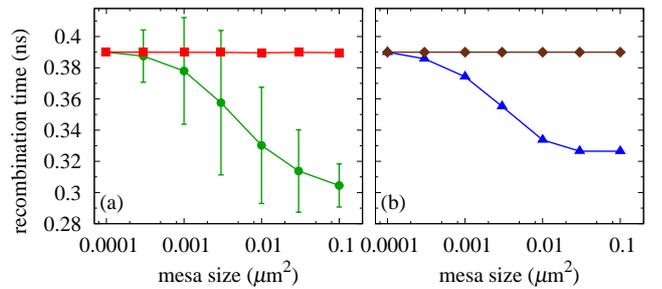}
\end{center}
\vspace{-5mm}\caption{\label{fig:area}(a) Dependence of the average exciton
  life time on the ``mesa'' size used in the simulation for an
  ensemble of dipole coupled dots (red squares) and SR-coupled
  dots (green circles). The error bars show formally the values of the standard
  deviation of the results obtained for various QD distributions;
  however, the actual distribution 
  of results is strongly asymmetric with a tail towards shorter life
  times and no results above the single dot exciton life time of
  0.39~ns. (b) Average exciton life time as a function of  the
  mesa size for two hypothetical QD ensembles: with dipole
  forces scaled up by a factor of 400 (blue triangles) and
  with SR couplings but emitting to separate reservoirs (brown
  diamonds). The results are averaged over a number of repetitions
  ranging from $10^{4}$ for the small systems to 100 for the largest
  system studied.}  
\end{figure}

As pointed out in Ref.~\cite{scheibner07}, a fingerprint of collective
spontaneous emission is the dependence of the exciton decay rate
on the number of emitters. In Fig.~\ref{fig:area}(a), we show the
simulation results for this dependence in the case of a QD sample
coupled only by the dipole forces (red squares) and in the presence of
the SR coupling (green circles). As expected from the discussion
presented above, there is no noticeable decrease of the exciton
lifetime in the case of LR EM coupling. However,
including the SR coupling leads to a considerable
acceleration of the spontaneous emission to a degree comparable to
that observed in the experiment \cite{scheibner07}. A similar result
is obtained in the hypothetical case of artificially magnified dipole
couplings, as shown by blue triangles in Fig.~\ref{fig:area}(b). 

Interestingly, for both kinds of couplings, the effect of a variation in the QD
surface density is quite similar: We have checked that decreasing the density by a
factor of 4 reduces the increase of the
decay rate)for a $0.03\,\mu$m$^{2}$ mesa to 23\% and 25\% of the
original value for the SR
and enhanced LR coupling models, respectively. 
Increasing the density by the same factor enhances this collective
effect by similar factors of 3.3\% and  2.6\%, 
respectively. At the same time, even in a sample with increased
density, the normal dipole coupling is still not sufficient to induce a
noticeable enhancement of the spontaneous emission.

The results of simulations show that strong SR coupling
between neighboring QDs is crucial for the
enhancement of the spontaneous emission, while the LR dipole
coupling encompassing the whole sample is irrelevant. It must be
noted, however, that the collective nature of the interaction between
the QDs and the EM field is still essential for the
observed effect. As follows from Eq.~\eqref{eq-motion}, collective
EM coupling affects the system dynamics in two ways: It
not only mediates the dipole interactions described by the coupling
constants $\Omega_{\alpha\beta}^{(\mathrm{rad})}$ but also leads to
the appearance of 
interference terms $\Gamma_{\alpha\beta}$ for $\alpha\neq \beta$ in
the dissipative part. These terms are absent in the hypothetical case
of QDs emitting to separate reservoirs (but still coupled by
interactions) \cite{sitek09b}. The brown diamonds in
Fig.~\ref{fig:area} show the exciton life times for such an artificial
system of QDs with additional SR interactions and coupled to separate
reservoirs. Clearly, 
no enhancement of the spontaneous emission is observed. 

\begin{figure}[tb]
\begin{center}
\includegraphics[width=85mm]{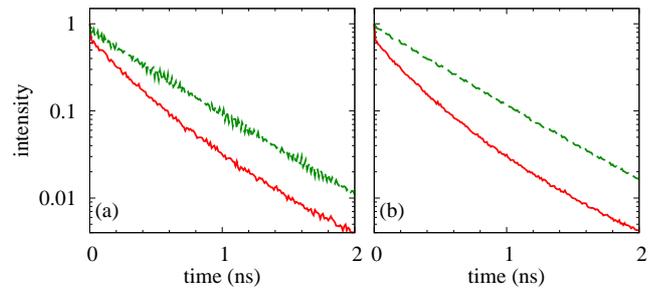}
\end{center}
\vspace{-5mm}\caption{\label{fig:filter}Time dependence of the luminescence
  intensity (normalized to the initial value) for detection at the
  photoluminescence maximum (solid red 
  lines) and 40~meV away from the maximum (dashed green lines): (a)
  for a $0.01\,\mu$m$^{2}$ mesa; (b) for a $0.1\,\mu$m$^{2}$ mesa. The
results are averaged over 100 different QD distributions.} 
\end{figure}

Another experimental fingerprint of collective emission
is the dependence of the luminescence decay time on the detection
energy  \cite{scheibner07}: Since in the spectral range around the
maximum of the ensemble photoluminescence (PL) more dots
contribute to the emission the collective effects in this energy
sector are stronger and the decay is faster. Our simulations,
shown in Fig.~\ref{fig:filter}, show that also 
this feature is reproduced in a model with SR coupling. Here,
we have simulated the measured signal from an ensemble with
$\sigma=28$~meV, using a Gaussian spectral filter 
$f(\omega)\sim\exp\{-(1/2)[(\omega-\omega_{0})/\Delta\omega]^{2}\}$
with $\hbar\Delta\omega=3$~meV, and $\hbar\omega_{0}$
at the PL maximum or 40~meV
away, for two sizes of the simulated mesas. As can be seen, in both
cases the PL decay at the PL maximum is indeed faster than in the tail
of the inhomogeneous distribution.

Our simulations of spontaneous emission from ensembles of dipole- and
SR-coupled QDs show that the emission rate in such a system can
indeed be increased due to collective coupling of the emitters to the
EM field. However, in view of relatively large
inhomogeneity of the QD transition energies, sufficiently strong
coupling between the dots is needed to stabilize the collective nature
of the emission. For typical inter-dot separations, fundamental dipole
interactions are too weak to play an important role in the emission 
kinetics. However, the presence of SR interactions, which may be
due to some kind of electronic coupling between the dots in
combination with Coulomb correlations,
leads to enhanced emission 
in quantitative agreement with the experimental results
\cite{scheibner07}. While such a coupling is likely to exist in QD
samples  \cite{kazimierczuk09}, its exact microscopic nature is
neither important 
for the emission dynamics nor can be inferred from it.
In any case, the presence of collective effects in the emission means that QDs are
not necessarily independent emitters, as usually assumed when
modeling their optical properties. Whether the increase 
of emission rate at low excitation is a signature of the true
superradiance that could result in a delayed outburst of radiation
under strongly inverted initial conditions \cite{skribanovitz73}
remains an open question. Also the evolution 
of inter-dot coherence in the process of
carrier trapping and relaxation, which would account for the
experimentally observed differences between emission under resonant
and non-resonant conditions, remains an interesting problem for further study.
Finally, in the course of further research
it may me interesting to clarify  whether this kind of cooperative
behavior can be responsible for radiative transfer of excitation
between remote dots \cite{sales04}.

\begin{acknowledgments}
This work was supported by the Foundation for
Polish Science under the TEAM programme, co-financed by the European
Regional Development Fund. Calculations have been partly carried out in
Wroclaw Centre for Networking and Supercomputing
(http://www.wcss.wroc.pl), grant No. 203.
\end{acknowledgments}

%\bibliographystyle{prsty}
%\bibliography{abbr,quantum}

\end{document}